\documentclass[
aps,prl
 amsmath,amssymb,
 reprint,%
]{revtex4-1}
\usepackage{chngcntr} 
\usepackage{etoolbox} 
\usepackage{graphicx}
\usepackage{float} 
\usepackage[mathlines]{lineno}
\usepackage[font=small]{caption,subcaption}
\usepackage{makecell}
\usepackage{tabularx}
\usepackage{dcolumn}
\usepackage{bm}
\usepackage{tikz}
\makeatletter
\@ifpackageloaded{float}{%
  \AtBeginEnvironment{figure}{\nolinenumbers}%
  \AtBeginEnvironment{figure*}{\nolinenumbers}%
  \AtBeginEnvironment{table}{\nolinenumbers}%
  \AtBeginEnvironment{table*}{\nolinenumbers}%
}{}
\makeatother
\usepackage{mathptmx}
\usepackage{color}
\usepackage[colorlinks,linkcolor=blue]{hyperref}
\usepackage{physics}
\usepackage{ulem}

\newcolumntype{Y}{>{\centering\arraybackslash}X}

\begin{document}

\title{A Novel Numerical Algorithms Optimization Method with Machine Learning Frameworks: Application on Real-time Plasmas Equilibrium Reconstruction in EXL-50U Spherical Torus}

\author{
    G.H. Zheng$^{1,4}$, S.F. Liu$^{1*,2}$, X. Gu$^{3\dagger}$, Y.P. Zhang$^{3}$, J. Li$^{3}$, Y. Liu$^{3}$, X.C. Lun$^{3}$, L. Xing$^{3}$, J.G. Chen$^{3}$,\\
    Z.Y. Chen$^{3}$, Y. Yu$^{3}$, D. Guo$^{3}$, Z.Y. Yang$^{4}$, H.S. Xie$^{3}$, X.M. Song$^{3}$, Y.J. Shi$^{3}$ and EXL-50U Team\\
    \small{$^1$ School of Physics, Nankai University, Tianjin 300071, People's Republic of China}\\
    \small{$^2$ Academy for Advanced Interdisciplinary Studies, Nankai University, Tianjin 300071, People's Republic of China}\\
    \small{$^3$ ENN Science and Technology Development Co., Ltd., Langfang 065001, People's Republic of China}  \\
    \small{$^4$ Southwestern Institute of Physics, Chengdu 610041, People's Republic of China}\\
    \small{$^*$Email : lsfnku@nankai.edu.cn}\\
    \small{$^\dagger$Email : guxiangc@enn.cn}
}

\begin{abstract}
    This work proposes for the first time a novel optimization method for numerical algorithms, which takes advantages of machine learning frameworks PyTorch and TensorRT, leveraging their modularity, low development threshold, and automatic tuning characteristics to achieve a real-time plasmas reconstruction algorithm called PTEFIT as an application in tokamak-based controlled fusion that combines performance, flexibility, and usability. The algorithm has been deployed and routinely operated on the EXL-50U spherical tokamak, with an average inference time of only 0.268ms per time slice at $129\times 129$ resolution, and has successfully driven feedback control of the maximum radial position of plasmas and isoflux control. We believe that its design philosophy has sufficient potential to accelerate development and optimization in GPU parallel computing, and is expected to be extended to other numerical algorithms.

    \quad

\end{abstract}

\maketitle

\section{Introduction}\label{sec:introduction}
Traditional numerical computation based on the first principle method always leads to unacceptable time complexity. Although with the rapid development of computation hardwares, it is a feasible way to accelerate relevant algorithms in parallel on GPUs, common researchers still struggle against the high threshold in complex optimization strategies on GPU programming.

In this case, machine learning is currently considered as the surrogates for high-fidelity, physical based algorithms. Trained machine learning models are almost entirely composed of matrix multiplication operations, whose computational speed can be significantly improved by parallel optimization. Furthermore, current machine learning frameworks have mature optimizations for various operators and easy to use, some of them even support automatic optimization of computational graphs and kernel functions to achieve maximum instruction throughput, cache hit rate, memory access efficiency, and thread utilization. However, although surrogate models often demonstrate computational speeds that meet requirements in most tasks, their black-box characteristics and limited generalization capability seriously affect their reliability and precision in solving scientific problems, and thus cannot yet serve as a mature solution to completely replace traditional numerical algorithms.

In fact, from a computational perspective, there is no essential difference between machine learning and numerical simulation. Therefore, we note that the machine learning frameworks' powerful optimization capabilities and low development threshold are expected to significantly improve the efficiency of numerical computation. As an application, this paper proposes $\mathbf{P}$yTorch-$\mathbf{T}$ensorRT-EFIT (PTEFIT), a GPU-parallel numerical algorithm in plasmas equilibrium reconstruction algorithm based on physical principles, constructed and inferred using machine learning frameworks PyTorch\cite{ansel_pytorch_2024} and TensorRT\cite{noauthor_nvidia_nodate}. PTEFIT takes advantages of PyTorch's modular characteristic in algorithm construction, which enables the algorithm to flexibly handle different configurations encountered in real-time; the constructed PyTorch modules can be automatically converted to TensorRT engine through Open Neural Network Exchange (ONNX)\cite{noauthor_github_nodate}. The automatic optimization during TensorRT engine building can completely replace manual optimization of GPU parallel computing programs. Ultimately, PTEFIT achieves a computation time of only approximately 0.268 ms for a single time slice at $129\times 129$ resolution, $\sim 10^4\times$ faster than offline EFIT and about $2\sim 10\times$ faster than traditional GPU-accelerated reconstruction.

For tokamak-based controlled fusion, equilibrium reconstruction refers to the algorithm that solves for the internal state and distribution of plasmas using measurement data provided in real-time by diagnostics under the axisymmetric approximation. The basic principle of equilibrium reconstruction is to solve the Grad-Shafranov (G-S) equation\cite{grad_hydromagnetic_1958,xie_what_2026}

\begin{equation}
    \Delta^*\Psi = -\mu_0RJ_\varphi = -\mu_0R^2\dfrac{\text dP}{\text d\Psi}
    - \mu_0^2F\dfrac{\text dF}{\text d\Psi}
    \label{eq:gs}
\end{equation}
\noindent to obtain the poloidal magnetic flux distribution $\Psi$ from the toroidal current density $J_\varphi$. In the equation, the Laplacian operator $\Delta^* = R^2\nabla\cdot(\nabla/R^2)$, $R$ denotes the radial distance from the symmetry axis, $P$ and $F$ represent the plasmas pressure and poloidal current, respectively. In experiments, $J_\varphi$ is typically pre-divided into several current filaments or parameterized by preset distributions, such that a finite number of quantities can determine its distribution, thereby enabling the solution through constructing a least squares system of equations

\begin{equation}
    G_\text{P}J_\varphi = b - G_\text{C}I_\text{C},
    \label{eq:lsq}
\end{equation}

\noindent where $G_\text{P}$ is the mutual inductance matrix between the reconstruction region and diagnostic devices; $b$ represents the physical quantities provided by external diagnostics, such as loop flux and poloidal magnetic field; $I_\text{C}$ and $G_\text{C}$ denote the poloidal field coil currents and their mutual inductance matrices with the diagnostics, respectively. Other quantities such as magnetic field, safety factor, inductance, and beta can be then calculated given $\Psi$ and $J_\varphi$.

Equilibrium reconstruction is crucial for tokamak experiments, as it can not only be used to analyze discharge quality and guide optimization strategies for subsequent experiments, but also provide real-time plasmas state during discharge for various feedback controls. EFIT (Equilibrium FITting code) is currently the most popular tokamak equilibrium reconstruction algorithm\cite{lao_reconstruction_1985}, and has been widely used in devices such as DIII-D, EAST, START, HL-2A, HL-3, NSTX, KSTAR, C-MOD, JET, MAST, QUEST, and EXL-50U\cite{berkery_kinetic_2021,lian_east_2017,li_kinetic_2013,zhang_reconstruction_2025,lao_equilibrium_1990,obrien_equilibrium_1992,lao_ll_mhd_2005,sabbagh_equilibrium_2001,jinping_equilibrium_2009,park_kstar_2011,jiang_kinetic_2021,appel_equilibrium_2001,in_resistive_2000,zwingmann_equilibrium_2003,li_yg_efit_2011,hongda_study_2006,xue_equilibrium_2019,appel_lc_unified_2006}. EFIT solves the G-S equation by repeatedly performing Picard iterations for each time slice, calculating relevant plasmas parameters after convergence is achieved. Although EFIT can adequately handle offline reconstruction tasks, its computation time for a single time slice is on the order of seconds, making it unsuitable for direct use in real-time feedback control during discharge.

Therefore, equilibrium reconstruction is a typical example for those traditional numerical algorithms facing the trouble to balance the computation accuracy and efficiency. There have been several works attempting to optimize the algorithm for improved computational speed, including methods such as reducing grid resolution\cite{moret_tokamak_2015}, presetting certain parameters in a specified tokamak that would otherwise be computed in real-time, and implementing parallel computation on GPUs\cite{huang_fast_2017,huang_gpu-optimized_2020,ma_acceleration_2018}. However, each of the aforementioned methods has its own limitations: reducing resolution degrades accuracy to some extent; frozening certain parameters lowers the algorithm's generality across different devices and configurations; traditional GPU programming requires complex manual tuning to achieve optimal performance, and the optimal configuration also varies with grid resolution, hardware devices, and other factors.

In addition to the above works, many studies have also attempted to surrogate equilibrium reconstruction algorithms with deep neural networks\cite{bonotto_reconstruction_2024,mcclenaghan_augmenting_2024,lu_fast_2023,joung_deep_2019,lao_application_2022,madireddy_efit-prime_2024,mitrishkin_new_2021,joung_gs-deepnet_2023,wai_neural_2022,wan_machine-learning-based_2023,zheng_real-time_2024,zheng_efit-mini_2025,zhou_physics-informed_2025,ding_physics-informed_2025,fiorenza_caronte_2025}, but still cannot be applied widely in experiments for its inherent limitations mentioned above.

In this case, PTEFIT is developed to satisfy the requirements of real-time tokamak plasmas control in accuracy and efficiency, while also having the capability to handle different flux surface configurations and the potential for cross-device migration. Currently, PTEFIT has been deployed on the EXL-50U spherical tokamak, successfully driving feedback control of the maximum radial position of plasmas and isoflux control, completely meeting the desired performance in experiments, and is general enough to be applied to other algorithms.

In addition to machine learning frameworks-based development, PTEFIT also possesses the following two major features:

$\bullet$ PTEFIT is entirely constructed based on physical principles, thus free from the constraints of black-box and generalization performance of machine learning methods. Compared to other numerical real-time reconstruction works mentioned above, PTEFIT can not only achieve the speed required for real-time operation without reducing resolution, but flexibly adapt to different flux surface configurations as well, and has a lower development and usage threshold.

$\bullet$ PTEFIT adopts algorithmic implementations completely consistent with EFIT wherever possible without affecting real-time performance, and its reconstruction results have been benchmarked against those of EFIT.

The remainder of this paper is organized as follows: Section \ref{sec:methods} introduces the relevant implementation details of PTEFIT; section \ref{sec:results} evaluates the computational efficiency of PTEFIT and its consistency with EFIT under advanced divertor configurations. Additionally, this section demonstrates the effects of feedback control over the maximum horizontal position of plasmas and isoflux control with PTEFIT participation; finally, section \ref{sec:summary} summarizes our work and provides an outlook for future development.

PTEFIT follows the tokamak coordinate system convention $COCOS = 7$\cite{sauter_tokamak_2013}, and all performance evaluations in this paper were conducted on an NVIDIA(R) RTX 4090 GPU device.

\section{Methods}\label{sec:methods}

PTEFIT is modularly constructed utilizing PyTorch features, comprising 21 computational modules, each of which can operate independently. The primary roles for these modules in reconstruction include generating Green's function matrices, determining current density parameters, solving G-S equations, calculating flux surface averaged quantities, etc. The section mainly introduces the implementation principles and optimization methods of them.

\subsection{Green's Function Matrices Generation}\label{subsec:greens}
In equilibrium reconstruction, the Green's function matrices relate the toroidal current to the poloidal field (magnetic field or flux) it produces. In a coordinate system composed of $R, Z$ in the poloidal plane, the contributions to the poloidal magnetic field $B$ and flux $\psi$ at $(r', z')$ from a toroidal current filament intersecting the poloidal plane at $(r, z)$ are respectively

\begin{equation}
    B_r = \dfrac{\mu_0I}{4\pi}\int_0^{2\pi}
    \dfrac{r'(z-z')\cos\varphi\text{d}\varphi}
    {\sqrt{(z-z')^2+(r-r'\cos\varphi)^2+(r'\sin\varphi)^2}^3},
\end{equation}

\begin{equation}
    B_z = \dfrac{\mu_0I}{4\pi}\int_0^{2\pi}
    \dfrac{(r^{\prime2}-rr'\cos\varphi)\text{d}\varphi}
    {\sqrt{(z-z')^2+(r-r'\cos\varphi)^2+(r'\sin\varphi)^2}^3},
\end{equation}

\noindent and

\begin{equation}
    \psi = \dfrac{\mu_0I}{4\pi}\int_0^{2\pi}
    \dfrac{rr'\cos\varphi\text{d}\varphi}
    {\sqrt{(z-z')^2+(r-r'\cos\varphi)^2+(r'\sin\varphi)^2}},
\end{equation}

\noindent where $I$ is the current carried by the filament and $\varphi$ represents the toroidal angle. The entire calculation involves only element-wise and reduce sum operations as long as computing the elliptic integrals by cumulative summation, which can be transformed into highly parallel matrix multiplications. Optimized by PyTorch's native GPU acceleration based on the CUDA backend, PTEFIT generates all Green's function matrices in only approximately 15s at $129\times 129$ resolution, while EFIT requires about 15 minutes. The Green's function matrices can be pre-generated and have no real-time requirements as they are only related to presetting parameters such as tokamak geometry and reconstruction region.

\subsection{Response Matrix and Least Squares Solution}\label{subsec:lsq}
The plasmas toroidal current density $J_\varphi$ can always be written as the direct product of a matrix $\mathbf{J}_\text{b}$ and undetermined coefficients $X$ whether by the current filament division or the parameterization. Substituting into equation \ref{eq:lsq} yields

\begin{equation}
    \mathfrak{R} X = b - G_\text{C}I_\text{C},
\end{equation}

\noindent which can also be expressed as

\begin{equation}
    \left[\begin{array}{c}\mathfrak{R} \\ G_\text{C}\end{array}\right]
    \left[\begin{array}{c}X \\ I_\text{C}'\end{array}\right]
    = \left[\begin{array}{c}b \\ I_\text{C}\end{array}\right]
\end{equation}

\noindent while considering the coil currents as free parameters, from which $\mathfrak{R} = G_\text{P}\mathbf{J}_\text{b}$ is the response matrix and $I_\text{C}'$ is the inverted coil current, respectively. Compared to the measured coil currents $I_\text{C}$, $I_\text{C}'$ is obtained through joint optimization with other diagnostics, which can not only equivalently account for the effects of vacuum vessel induced currents to some extent, but also mitigate the vertical displacement of plasmas caused by Picard iteration. The construction of $\mathfrak{R}$ involves only trivial matrix multiplications and can therefore be efficiently completed by GPU.

In PTEFIT, $J_\varphi$ is parameterized in polynomial by assuming the forms of $P'$ and $FF'$ within the plasmas boundary are

\begin{equation}
    P' = \sum_{i=0}^{N_{P'}}a_i(\tilde\Psi^i - \tilde\Psi^{N_{P'}})
\end{equation}

\noindent and

\begin{equation}
    FF' = \sum_{j=0}^{N_{FF'}}b_j(\tilde\Psi^j - \tilde\Psi^{N_{FF'}}),
\end{equation}

\noindent which is consistent with EFIT. From equation \ref{eq:gs}, $\mathbf{J}_\text{b}$ and $X$ can be expressed as

\begin{align}
    \mathbf{J}_\text{b} = 
    \text{\Large (} & R(1 - \tilde\Psi^{N_{P'}}), R(\tilde\Psi - \tilde\Psi^{N_{P'}}), ..., \notag \\
    \phantom{(} & \dfrac{\mu_0}{R}(1 - \tilde\Psi^{N_{FF'}}), ..., \dfrac{\mu_0}{R}(\tilde\Psi^{N_{FF'} - 1} - \tilde\Psi^{N_{FF'}})\text{\Large )}
\end{align}

\noindent and

\begin{equation}
    X = (a_0, a_1, ..., a_{N_{P'}}, b_0, b_1, ..., b_{N_{FF'}})^\text{T},
\end{equation}

\noindent respectively, where $\tilde\Psi$ is the normalized flux, $N_{\text{P}'}$ and $N_{\text{FF'}}$ are the maximum polynomial orders of $P'$ and $FF'$, respectively. Therefore, solving for $J_\varphi$ requires first calculating the fluxes at the axis and boundary, as well as a boolean matrix representing the plasmas existence region, which will be discussed in section \ref{subsec:ox}.

PTEFIT provides two methods for solving the least squares equations: constructing normal equations and QR decomposition by Gram-Schmidt orthogonalization. One should be pointed that PTEFIT does not employ the singular value decomposition (SVD) method used in EFIT, because the rounding errors induced by SVD is unacceptable in single-precision floating point number, which is generally operated in GPU. It can be shown that the QR decomposition based on Gram-Schmidt orthogonalization is numerically stable\cite{bjorck_solving_1967}, but requires longer computation time; constructing and solving normal equations is faster but causes the increasing condition number of the coefficient matrices, making it suitable for cases where the dimension of $X$ is relatively low.

\subsection{Solution of G-S Equation}\label{subsec:gs}
The G-S equation can be transformed through finite differences into a $(N_R - 2) \times (N_Z - 2)$ dimensional pentadiagonal matrix equation with Dirichlet boundary conditions, where $N_R$ and $N_Z$ represent the number of grid points in the $R$ and $Z$ directions, respectively. Directly solving the equation incurs enormous computational overhead, but it can be transformed into $N_Z - 2$ tridiagonal matrix equations to solve in parallel, where the coefficient matrices can be pre-computed. Detailed principles can be found in reference\cite{huang_gpu-optimized_2020}.

\subsection{Processing on Flux Surface}\label{subsec:ox}
The flux surfaces solved from the G-S equation must be processed to obtain the normalized flux $\tilde\Psi$ and the plasmas existence region for constructing $\mathbf{J}_\text{b}$, as mentioned in section \ref{subsec:lsq} above.

The normalized flux distribution can be derived by computing flux values at the magnetic axis and boundary of the plasmas. For divertor configurations, the axis and boundary correspond to an O-point and an X-point in the flux distribution, both of them are extremal points satisfying $\nabla\Psi = 0$. Furthermore, the Hessian matrix $\mathbf{H}$ at O-points and X-points satisfies $\det\mathbf H > 0$ and $\det\mathbf H < 0$, respectively. Notably, it is impossible to find grid points where $\nabla\Psi$ is exactly $0$ in a discretized flux distribution, and alternative schemes based on sign comparison of finite difference products may fail due to non-uniform flux variation in different directions near X-points. To address this, a second-order polynomial fitting, which formally allows for the existence of O-points and X-points, is utilized in the vicinity of each grid point, enabling exact analytical solutions where $\nabla\Psi = 0$. Meanwhile, under the parameterized distribution, the Hessian matrix $\mathbf H$ has a simple form and is easy to compute.

Polynomial fitting requires solving $(N_R - 2)\times (N_Z - 2)$ least squares equations, but since the grid point distribution is fixed, the algorithm only needs to perform matrix multiplications for this step in real-time by pre-constructing normal equations and inverting the coefficient matrices.

The flux varies monotonically from the magnetic axis to the boundary within the plasmas region, but exhibits complex behavior outside the last closed flux surface (LCFS). Therefore, the plasmas existence region must be identified first to find the axis and boundary from all O- and X-points. PTEFIT employs the same principle as described in reference\cite{moret_tokamak_2015} to determine the plasmas region, but executes in parallel on the GPU.

For limiter configurations, the plasmas boundary is not located at an X-point. However, for any single plasma blob, the boundary flux can be determined by comparing the flux values of all limiter vertices and X-points within the plasmas region.

For doublet and droplet configurations, only the flux at limiter positions is compared when selecting the boundary, which can be achieved by appending a value with very large absolute magnitude and sign opposite to $I_\text{P}$ to X-point fluxes during comparison.

\subsection{Flux Surface Averaged Quantities}\label{subsec:surfs}
Flux surface averaged quantities, including the safety factor, are crucial for plasmas transport and confinement performance, and therefore need to be computed rapidly for subsequent real-time control requirements. A flux surface averaged quantity $Q$ can be expressed as a specific integral form along a closed flux surface:

\begin{equation}
    Q_{\mu, \nu} = \oint_l |\nabla\Psi|^\mu R^\nu \text dl,
\end{equation}

\noindent where $l$ is the poloidal path along a certain closed flux surface, and $\mu, \nu$ are arbitrary integers. In particular, the safety factor can be expressed as $q = \dfrac{\mu_0F}{2\pi}Q_{-1, -2}$. The only difficulty in computing $Q_{\mu, \nu}$ lies in obtaining the poloidal path $l$ of the closed flux surface.

\begin{figure}
    \centering
    \includegraphics[width=.85\linewidth]{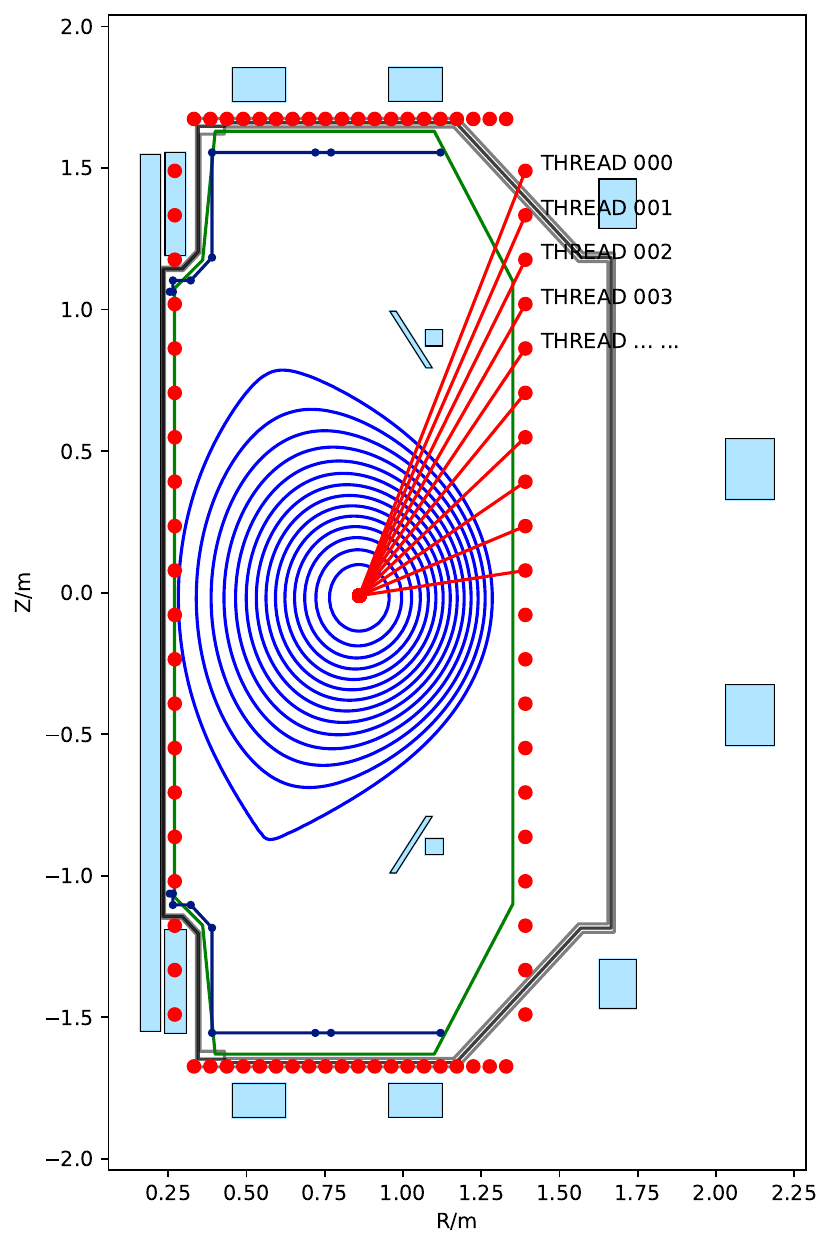}
    \caption{Schematic diagram of the bisection method for finding flux surface vertices. The flux outside the plasmas region need to be filtered first. For each thread, the initial endpoints are located at the magnetic axis and the reconstructed boundary, respectively, and the points on the reconstructed boundary must be arranged continuously in clockwise or counterclockwise order.}
    \label{fig:surfs}
\end{figure}

PTEFIT adopts the bisection method to determine flux surface vertices. By filtering the flux outside the plasmas region using the boolean matrix obtained in section \ref{subsec:ox}, the flux monotonicity required by the bisection method is guaranteed. As shown in figure \ref{fig:surfs}, the bisection method can efficiently take advantages of GPU parallel characteristic, with each thread processing one chord. The initial endpoints are taken at the magnetic axis and a point on the reconstructed boundary, respectively. After each search, linear interpolation is performed to obtain the new vertex flux. It should be noted that the initial vertices taken on the reconstructed boundary must be arranged continuously in clockwise/counterclockwise order to ensure the flux surface element $\text dl$ can be correctly computed subsequently. Under typical device scales, only about 10 iterations are needed to reduce the flux surface error to approximately 1mm, which is sufficient considering the error impact from linear interpolation. After obtaining the flux surface vertices, the subsequent geometric parameters of the plasmas (minor radius, elongation, etc.) and global parameters (beta, internal inductance, stored energy) can also be derived through simple calculations.

PTEFIT can also search for flux surfaces under doublet or droplet configurations through appropriate selection of iteration initial endpoints and additional filtering operations. Details are not elaborated in this article.

\subsection{Optimization}\label{subsec:build}

In addition to parallelization, the main optimization techniques employed by PTEFIT during algorithm construction are as follows:

1. PTEFIT is based on the slow flux variation assumption, performing only one Picard iteration per time slice during a continuous discharge.

2. To ensure that the algorithm can be fully captured as a CUDA graph, boolean variables and multiply-add operations are used instead of conditional statements to avoid potential graph non-determinism. Operators that cannot be captured as CUDA graphs have been rebuilt to TensorRT custom plugins to explicitly define their behavior.

3. All data that can be pre-computed before discharge are computed in advance and frozen in real-time to improve efficiency.

4. The entire computation flow is defined to execute on the GPU to reduce data transfer latency, from which division and square root operations are avoided wherever possible. Memory, cache, and thread optimizations are automatically completed by TensorRT.

\section{Results}\label{sec:results}

\subsection{Consistency with Offline EFIT}\label{subsec:consist}
\begin{figure}[h]
    \centering
    \includegraphics[width=\linewidth]{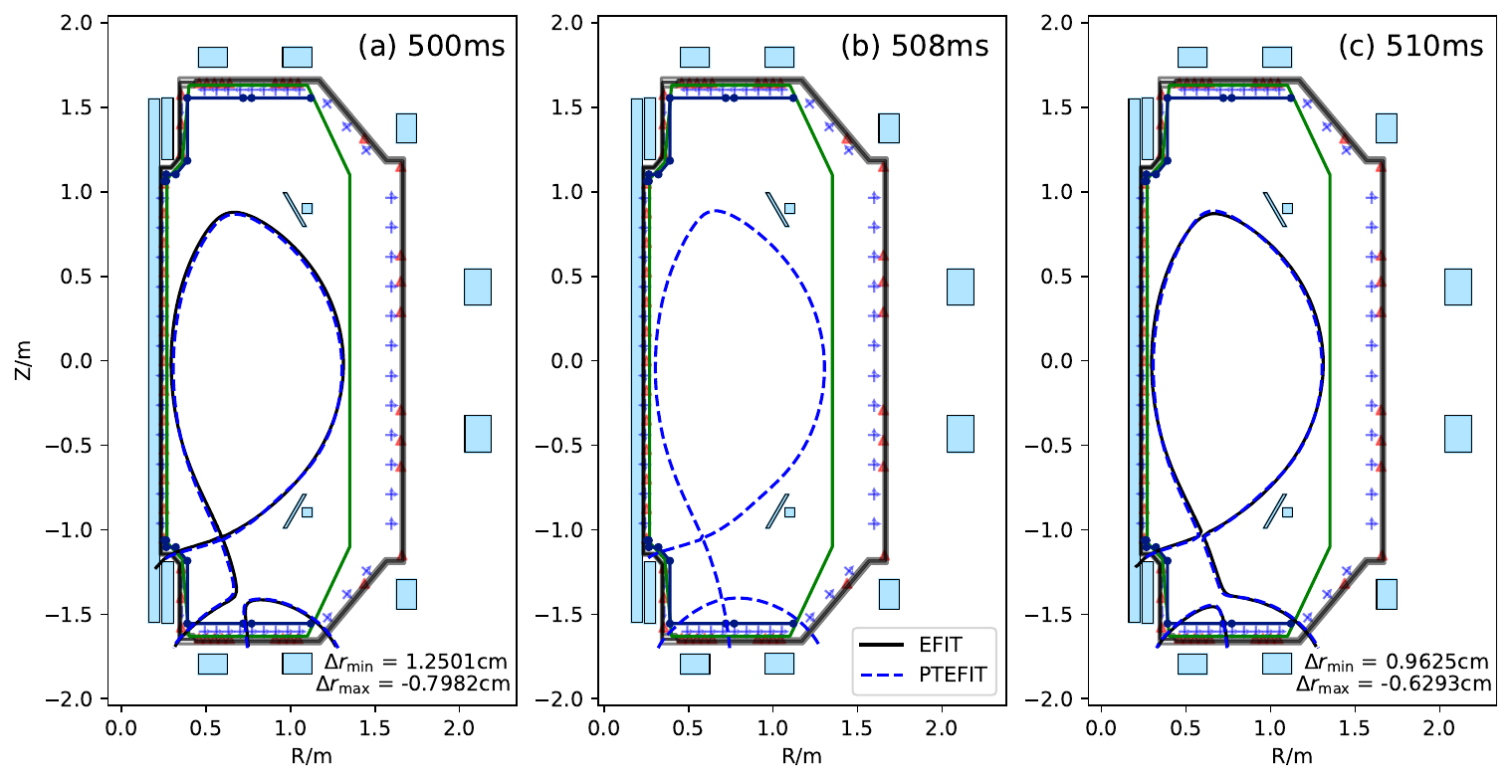}
    \caption{Comparison of LCFS results obtained by PTEFIT and offline EFIT for XPT configuration reconstruction of shot \#13899. Here, $\Delta r_{\min}$ and $\Delta r_{\max}$ are the deviations between the two at the minimum and maximum radial positions of the LCFS, respectively.}
    \label{fig:xpt}
\end{figure}

We selected the XPT (X-Point Target) configuration recently successfully achieved on the EXL-50U spherical tokamak\cite{shi_strategy_2025,shi_achievement_2025} to compare the consistency between PTEFIT and offline EFIT. Figures \ref{fig:xpt}(a)(c) compare the LCFS obtained by PTEFIT and offline EFIT for two time slices in shot \#13899, from which PTEFIT and EFIT show high consistency in both X-divertor minus and X-divertor plus configurations, with the LCFS reconstruction deviation controlled to approximately 1cm. Notably, due to the computational speed limitations, inter-shot offline EFIT can only operate reconstruction at a time slice period of 10ms, potentially missing certain important experimental phenomena; while PTEFIT computes in real-time at 1kHz resolution during discharge, enabling observation of critical moments such as the transition from X-divertor minus to X-divertor plus immediately after discharge (figure \ref{fig:xpt}(b)).

\subsection{Computation Efficiency}\label{subsec:speed}
\begin{figure}[h]
    \centering
    \includegraphics[width=.95\linewidth]{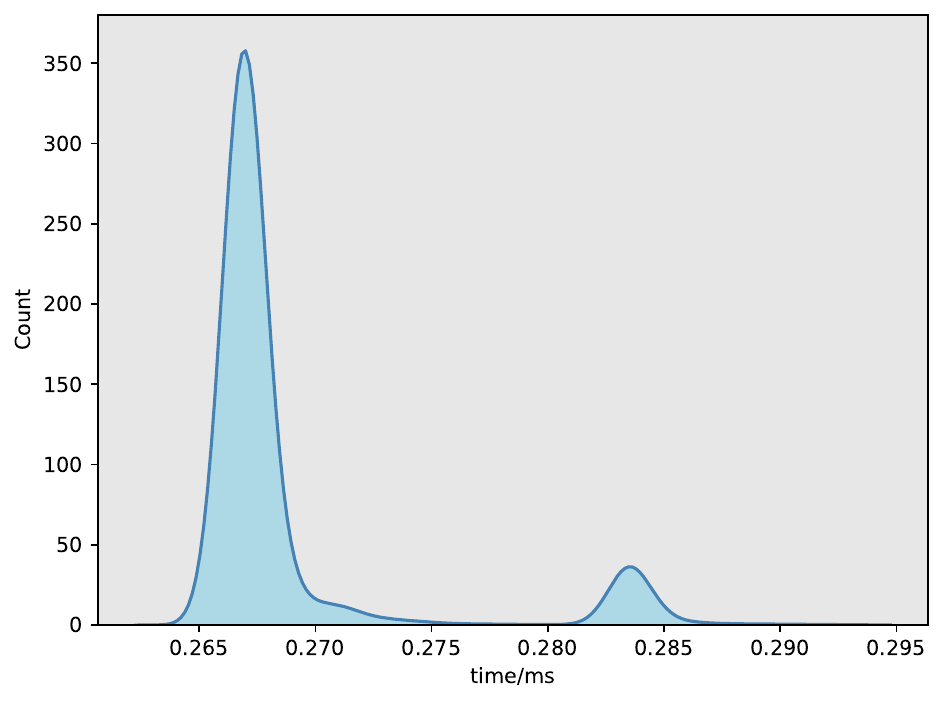}
    \caption{Distribution of PTEFIT computation time per time slice.}
    \label{fig:speed}
\end{figure}

PTEFIT has achieved sub-millisecond computational speed with 0.268ms average computation time per slice under the configuration of $129\times 129$ grid resolution, 33 profile points, and normal equation construction as the least squares driver. Figure \ref{fig:speed} shows the distribution of PTEFIT computation speed over 10,000 consecutive time slices.

\subsection{Feedback Control over $R_{\max}$}\label{subsec:rmax}
\begin{figure}
    \centering
    \includegraphics[width=.45\textwidth]{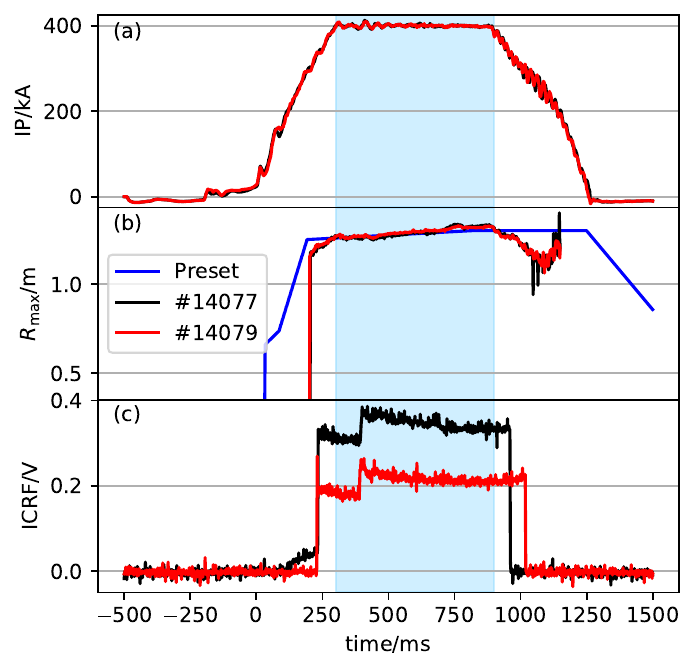}
    \caption{Demonstration of PID feedback control effects using the maximum radial position $R_{\max}$ of plasmas provided in real-time by PTEFIT for two shots. (a) Plasmas current waveform. (b) Preset $R_{\max}$ and $R_{\max}$ obtained from PTEFIT real-time reconstruction. (c) ICRF intensity waveform.}
    \label{fig:rmax}
\end{figure}

Maintaining the maximum radial position $R_{\max}$ of plasmas stable is crucial for ion cyclotron range of frequencies (ICRF) heating\cite{noterdaeme_interaction_1993}. During ICRF experiments on EXL-50U, PTEFIT provides $R_{\max}$ in real-time, enabling proportional-integral-derivative (PID) feedback control. Figure \ref{fig:rmax} shows two shots of discharge with $R_{\max}$ control driven by PTEFIT, where different intensities of ICRF heating were applied. As shown in figure \ref{fig:rmax}(b), during the current flat-top period, the PTEFIT-driven $R_{\max}$ feedback control can maintain the target deviation at approximately 2cm.

\subsection{Isoflux Control}\label{subsec:isoflux}
\begin{figure}[h]
    \centering
    \includegraphics[width=.48\textwidth]{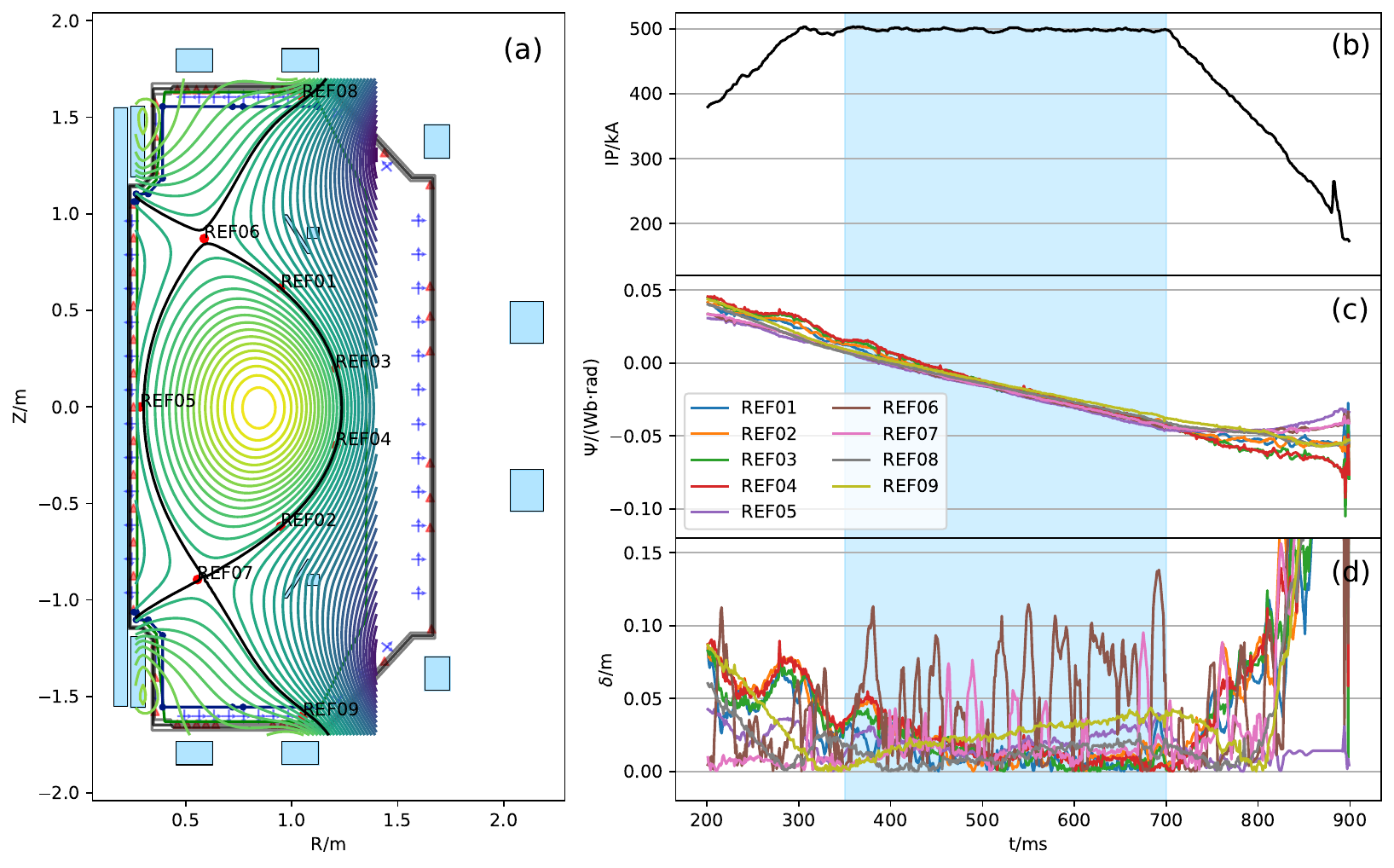}
    \caption{Schematic diagram of isoflux control for shot \#14488. (a) Flux surface distribution at 600 ms. (b) Plasmas current waveform. (c) Temporal evolution of flux at each reference point. (d) Temporal evolution of distance between each reference point and the LCFS. The background is light blue during the isoflux control period.}
    \label{fig:isoflux}
\end{figure}

Isoflux control refers to the multi-input multi-output (MIMO) feedback conrtol by observing the flux difference between LCFS and preset reference points. Figure \ref{fig:isoflux} shows the waveforms and isoflux control effects of EXL-50U shot \#14488. 

In this shot, isoflux control takes over from 350ms to 700ms. Among them, the red dots are the preset reference points, including 2 reference divertor strike points, 2 reference X-points, and 5 other reference points located on the LCFS. Figures \ref{fig:isoflux}(b)(c)(d) sequentially show the plasmas current, the flux at each reference point, and the distance from the LCFS. During isoflux control, it can be seen that the flux values at each reference point are nearly identical, and the distance between each reference point (except X-points) and the LCFS averages approximately 2cm. However, due to the influence of vertical displacement events (VDEs), only one of the two X-points can be located on the LCFS at the same time. VDEs grow so rapidly that isoflux control at a period of 1ms is not sufficient to suppress this instability. The improved future work combining vertical displacement control with isoflux control may be considered to mitigate the problem.

\section{Summary}\label{sec:summary}
Based on physical principles, we take advantages of machine learning frameworks to optimize numerical algorithms, implementing the real-time reconstruction algorithm PTEFIT, which is constructed in PyTorch and executes in real-time under TensorRT. PTEFIT has been successfully deployed on the EXL-50U spherical tokamak to support experiments including XPT configuration discharges, ICRF, and isoflux control. Leveraging the Python API and inherent modularity provided by PyTorch, the algorithm has an extremely low development and usage threshold, and is easy for migration; the automatic optimization and C++ API of TensorRT further reduce development difficulty while ensuring real-time performance and compatibility, ultimately achieving an average reconstruction time of 0.268 ms for single time slice.

PTEFIT maintains consistency with EFIT in both algorithm and principle wherever possible, and is benchmarked against EFIT results, with the reconstructed LCFS showing an average deviation of approximately 1cm from offline EFIT. In terms of real-time deployment, both the control targeting plasmas radial position and the isoflux maintain the accuracy of approximately 2cm.

Although PTEFIT's computational efficiency already meets experimental requirements, the results need to be transferred to CPU and then sent to the control system via reflective memory for feedback control, leaving room for improvement in the overall latency. Future work may consider integrating the feedback control algorithm into PTEFIT to directly output coil voltages for improved computational efficiency. Additionally, it is also meaningful to combine kinetic diagnostics with integrated modeling to calculate accurate profiles, which could be applied to profile feedback control and other experiments.

Finally, we believe that PTEFIT and its design philosophy, which taking fully advantages of mutual machine learning frameworks for numerical optimization, have sufficient generalization capabilities applicable to most traditional numerical algorithms efficiently.

\begin{acknowledgments}
    This work is supported by National Natural Science Foundation of China under Grant No.12275142 and No.12275354, National MCF Energy R\&D Program under Grant No.2024YFE03020001. Besides, we are particularly grateful to ENN for the scientific research funding supporting this study.
\end{acknowledgments}

\bibliographystyle{unsrt}
\bibliography{ref}

\end{document}